\documentclass[preprint,12pt]{elsarticle}

\usepackage[margin=0.9in]{geometry}
\usepackage{graphicx, xcolor}
\usepackage{amsmath,amssymb,amsfonts,amsthm,bm,mathtools,stmaryrd}
\usepackage{subfigure}
\usepackage{subcaption}
\usepackage{hyperref} 

\hypersetup{%
  colorlinks=true,
  linkcolor=blue,
  linkbordercolor=red,
  citecolor=red,
}

\journal{Communications in Nonlinear Science and Numerical Simulation}

\begin{document}

\begin{frontmatter}



\title{A thermodynamically consistent discretization of 1D thermal-fluid models using their metriplectic 4-bracket structure}


\author[inst1]{William Barham\corref{cor1}}
\cortext[cor1]{Corresponding author}
\ead{william.barham@utexas.edu}


\author[inst1]{Philip J. Morrison}

\author[inst2]{Azeddine Zaidni}

\affiliation[inst1]{organization={Department of Physics and Institute for Fusion Studies, The University of Texas at Austin},
            addressline={2515 Speedway},
            city={Austin},
            postcode={78712},
            state={Texas},
            country={United States}}

\affiliation[inst2]{organization={Mohammed VI Polytechnic University, College of Computing},
    addressline={Lot 660, Hay Moulay Rachid},
    city={Ben Guerir},
    postcode={43150},
    state={Marrakech-Safi},
    country={Morocco}}

\begin{abstract}
Thermodynamically consistent models in continuum physics, i.e.\ models which satisfy the first and second laws of thermodynamics, may be expressed using the metriplectic formalism. In this work, we leverage the structures underlying this modeling formalism to preserve thermodynamic consistency in discretizations of a fluid model. The procedure relies (1) on ensuring that the spatial semi-discretization retains certain symmetries and degeneracies of the Poisson and metriplectic 4-brackets, and (2) on the use of an appropriate energy conserving time-stepping method. The minimally simple yet nontrivial example of a one-dimensional thermal-fluid model is treated. It is found that preservation of the requisite symmetries and degeneracies of the 4-bracket is relatively simple to ensure in Galerkin spatial discretizations, suggesting a path forward for thermodynamically consistent discretizations of more complex fluid models using more specialized Galerkin methods. 
\end{abstract}



\begin{keyword}
Navier-Stokes-Fourier
\sep Structure-preserving discretization 
\sep Thermodynamic consistency
\sep Metriplectic dynamics
\sep Hamiltonian structure
\PACS 47.11.-j 
\sep 02.70.-c
\sep 02.60.-x
\MSC 76M10 \sep 37K05 \sep 37L65 \sep 80A17
\end{keyword}

\end{frontmatter}


\section{Introduction} \label{sec:intro}

The Navier-Stokes-Fourier system is known to satisfy the first and second laws of thermodynamics. Therefore, it is desirable that a numerical scheme for this system should likewise be consistent with the laws of thermodynamics. This work derives such a scheme in one spatial dimension by preserving the symmetries of the Poisson bracket and the metriplectic $4$-bracket formulation of the model in its spatial discretization. The Hamiltonian formalism is well established as a powerful modeling tool for ideal fluids \cite{RevModPhys.70.467}. A dissipative extension of Hamiltonian mechanics, known as the metriplectic formalism, has been known for several decades \cite{MORRISON1986410}, however the formalism has recently been extended \cite{PhysRevE.109.045202} to explicate, in an algorithmic fashion, the proper design of thermodynamically consistent models (see its application in deriving a thermodynamically consistent Cahn-Hilliard-Navier-Stokes model \cite{osti_2426411}). The novelty in this new formalism is an object known as the metriplectic $4$-bracket, which allows one to recover the previous notion of a metriplectic $2$-bracket with the required degeneracy for energy conservation. The metriplectic $2$-bracket has previously been used in the discretization of collisional kinetic plasmas \cite{10.1063/1.4998610, jeyakumar2024}, and the calculation of MHD equilibria \cite{bressan2018relaxation}. The formalism has also been employed in reduced order modeling \cite{gruber2023energetically}. Previous thermodynamically consistent discretizations of thermal-fluid models have been derived based on the Lagrange d'Alembert principle, a dissipative extension of Lagrangian mechanics \cite{doi:10.1142/S0218202524500027}. There is also prior work considering thermodynamic consistency in the context of sub-grid parameterizations in atmospheric modeling \cite{gassmann2015local}. This paper is the first to use a metriplectic $4$-bracket in the design of a numerical method. 

It is worth mentioning that the natural variables for fluid models using the metriplectic formulation use entropy rather than internal energy as a prognostic variable. This makes our formulation incompatible with many tools in common practice for numerical methods for hyperbolic conservation laws. Standard methods use conservation form to great effect in deriving finite difference \cite{richtmyer_morton_1967, toro2013riemann}, finite volume \cite{toro2013riemann, leveque_fv_2002}, and discontinuous Galerkin methods \cite{cockburn_shu_2001, hesthaven_warburton_2008}, with well-developed stabilization techniques for shocks inextricably connected to the use of conservation form. However, there is precedent in the literature for using a skew-symmetric split form---a weak formulation that incorporates both the advective and conservative forms of the transport operator---rather than solely using conservative form to simulate fluid models \cite{morinishi_1998, gassner_2013, gassner_winters_2014, palha_2017}. Similar to this work, the motivation for employing these split forms is to construct invariant-preserving schemes. Moreover, prior studies have explored the use of entropy, rather than total energy, as a prognostic variable in compressible flow simulations (see \cite{shakib1991new} and references therein). While this work builds on prior research, the approach proposed herein does not aim to compete with the state-of-the-art methods based on conservation form without substantial further research addressing the need for stabilization and shock-capturing techniques tailored to this formalism. 

A metriplectic model is prescribed by: a Hamiltonian, $H$, a functional of the dynamical fields; a Poisson bracket, $\{\cdot, \cdot\}$, a bilinear map on functionals of the fields; an entropy, $S$, a Casimir invariant of the Poisson bracket which generates the dissipative dynamics; and a metriplectic $4$-bracket, $(\cdot, \cdot; \cdot, \cdot)$, a $4$-linear map on the algebra of functionals. A Poisson bracket has the following properties: $\forall F,G,H$ and $\forall a,b \in \mathbb{R}$,
\begin{equation}
\begin{aligned}
    &\{F, aG + bH\} = a \{F,G\} + b \{F, H\} \,, \\
    &\{F, G\} = - \{G, F\} \,,
    \\
    &\{ F, \{G, H\} \} + \{G, \{H, F \} \} + \{ H, \{F, G\} \} = 0 \,, \\
    &\{F, G H \} = G \{F, H \} + \{F, G \} H \,.
\end{aligned}
\end{equation}
A Casimir invariant is a degeneracy of the Poisson bracket, i.e. \ a functional, $C$, such that $\{F, C\} = 0$ $\forall F$. As previously stated, the entropy, $S$, must be a Casimir invariant of the Poisson bracket in the metriplectic formalism. Finally, the metriplectic $4$-bracket is a $4$-linear map with the following properties: $\forall F,K,G,N$,
\begin{equation}
\begin{aligned}
    (F, K; G, N) &= - (K, F; G, N) \,,
    \\
    (F, K; G, N) &= (G, N; F, K) \,,
    \\
    (FH, K; G, N) &= F(H, K; G, N) + (F, K; G, N) H \,,
    \\
    (F, G; F, G) &\geq 0 \,.
\end{aligned}
\end{equation}
For any observable, $F$, its evolution is prescribed by $\dot{F} = \{F, H\} + (F, H; S, H)$. 

Thermodynamic consistency is guaranteed in the metriplectic formalism by the following properties: (i) the entropy is a Casimir invariant of the Poisson bracket, $\{F,S\} = 0$ $\forall F$; (ii) antisymmetry of the Poisson bracket: $\{F,G\} = - \{G,F\}$; (iii) antisymmetry of the $4$-bracket, $(F, K; G,N) = -(K, F; G, N)$; (iv) finally, non-negative entropy production is ensured by the semi-definiteness of the bracket: $(S, H; S, H) \geq 0$. Together, these ensure thermodynamic consistency:
\begin{equation}
    \dot{H} = \{H, H\} + (H, H; S, H) = 0 \,,
    \quad \text{and} \quad
    \dot{S} = \{S, H\} + (S, H; S, H) = (S, H; S, H) \geq 0. 
\end{equation}
The reader is directed to \cite{PhysRevE.109.045202, osti_2426411, zaidni2024metriplectic} for a more complete account of the metriplectic $4$-bracket formalism. A concrete realization of these abstract objects will be given subsequently.

\section{A thermal-fluid model and its metriplectic structure}

A thermodynamically-consistent model of compressible flow, frequently called the Navier-Stokes-Fourier system, was shown to possess metriplectic $4$-bracket structure \cite{osti_2426411}. In a single spatial dimension, the equations of motion are given by
\begin{equation} \label{eq:nsf_system}
\begin{aligned}
    &\partial_t \rho
    + \partial_x (\rho u) = 0 \,,
    \quad 
    \partial_t (\rho u) 
    +
    \partial_x (\rho u^2)
    + \partial_x p
    = \partial_x ( \mu \partial_x u ) \,,
    \\
    &\partial_t (\rho s)
    + 
    \partial_x(\rho s u)
    =
    \frac{\mu}{T} (\partial_x u)^2
    +
    \partial_x \left( \frac{\kappa}{T} \partial_x T \right) 
    + \frac{\kappa}{T^2} (\partial_x T)^2 \,,
\end{aligned}
\end{equation}
where $\mu$ and $\kappa$ are the viscosity and thermal-conductivity coefficients, respectively, and there exists an internal energy $U = U(\rho, s)$ such that the pressure and temperature are prescribed by $p = \rho^2 \partial_\rho U$ and $T = \partial_s U$. It may be shown that this model possesses a metriplectic structure.

It is convenient to use density coordinates: $(\rho, m, \sigma) = (\rho, \rho u, \rho s)$ when writing the metriplectic structure of the Navier-Stokes-Fourier system. The Hamiltonian is given by
\begin{equation}
    H[\rho, m, \sigma]
    =
    \int_\Omega
    \left(
    \frac{1}{2} \frac{m^2}{\rho}
    +
    \rho U \left(\rho, \frac{\sigma}{\rho} \right) 
    \right)
    \mathsf{d} x \,,
\end{equation}
and, using the functional derivative shorthand $\delta F/\delta u = F_u$, the Poisson bracket \cite{pjmG80} is given by
\begin{equation}
    \{F, G\}
    =
    - \int_\Omega
    \bigg[
    m \left(
    F_m \partial_x G_{m}
    -
    G_{m} \partial_x F_m
    \right)
    + 
    \rho \left(
    F_m \partial_x G_{\rho}
    -
    G_{m} \partial_x F_{\rho}
    \right) 
    +
    \sigma \left(
    F_m \partial_x G_{\sigma}
    -
    G_{m} \partial_x F_{\sigma}
    \right)
    \bigg]
    \mathsf{d} x \,.
\end{equation}
Assuming homogeneous or periodic boundary conditions, the evolution law $\dot{F} = \{F, H\}$ for arbitrary $F = F[\rho, m, \sigma]$ recovers the conservative part of the dynamics given by the right-hand side of equation \eqref{eq:nsf_system}. This Poisson bracket possesses a Casimir invariant of the form $S[\rho, m, \sigma] = \int_\Omega \sigma \mathsf{d} x$. This is the total entropy, and will be used as the generator for the dissipative dynamics. 

The metriplectic structure is prescribed by a $4$-bracket constructed using the Kulkarni-Nomizu product (see e.g. \cite{PhysRevE.109.045202, osti_2426411, zaidni2024metriplectic}) from the following symmetric operators:
\begin{equation} \label{eq:kn_components}
    M(F, G)
    =
    F_\sigma G_\sigma \,, 
    \quad \text{and} \quad 
    \Sigma(F, G)
    =
    (\partial_x F_m) \frac{\mu}{T} (\partial_x G_m) 
    +
    (\partial_x F_\sigma) \frac{\kappa}{T^2} (\partial_x G_\sigma) \,. 
\end{equation}
The Kulkarni-Nomizu product is given by
\begin{equation}
\begin{aligned}
    (\Sigma \varowedge M)(F, K, G, N)
    =
    \Sigma(F, G) M(K, N) 
    &- \Sigma(F, N) M(G, K) \\
    &+ M(F, G) \Sigma(K, N) 
    - M(F, N) \Sigma(G, K) \,,
\end{aligned}
\end{equation}
from which one then defines the $4$-bracket:
\begin{multline}
    (F, K; G, N) 
    =
    \int_\Omega (\Sigma \varowedge M)(F, K, G, N) \mathsf{d} x 
    \\
    =
    \int_\Omega
    \frac{1}{T}
    \bigg[
    \mu
    \left(
    K_{\sigma} \partial_x F_m
    -
    F_{\sigma} \partial_x K_{m}
    \right) 
    \left(
    N_{\sigma} \partial_x G_{m}
    -
    G_{\sigma} \partial_x N_{m}
    \right) \\
    +
    \frac{\kappa}{T}
    \left(
    K_{\sigma} \partial_x F_{\sigma}
    -
    F_{\sigma} \partial_x K_{\sigma}
    \right) 
    \left(
    N_{\sigma} \partial_x G_{\sigma}
    -
    G_{\sigma} \partial_x N_{\sigma}
    \right)
    \bigg]
    \mathsf{d} x \,.
\end{multline}
The rationale for choosing the operators $M$ and $\Sigma$ as given in \eqref{eq:kn_components} comes from a closer examination of the implied dissipative evolution:
\begin{equation} \label{eq:dissipative_flow}
\begin{aligned}
    (F, S)_H
    &=
    \int_\Omega
    (\Sigma \varowedge M)(F, K, G, N)
    \mathsf{d} x 
    =
    \int_\Omega
    \left(
    - \Sigma(F, H) M(S, H) 
    + M(F, S) \Sigma(H, H) 
    \right)
    \mathsf{d} x \\
    &=
    \int_\Omega
    \left(
    - 
    \left[
     (\partial_x F_m) \mu \partial_x u
    +
    (\partial_x F_\sigma) \frac{\kappa}{T} (\partial_x T)
    \right]
    + 
    F_\sigma
    \left[
    \frac{\mu}{T} (\partial_x u)^2 
    +
    \frac{\kappa}{T^2} (\partial_x T)^2
    \right]
    \right)
    \mathsf{d} x \,.
\end{aligned}
\end{equation}
By letting $M = F_\sigma G_\sigma$, we find that $\Sigma(H,H)$ is the entropy production rate while $\Sigma(F,H)$ gives rise to the reciprocal couplings which ensure energy conservation. This rationale for finding the metriplectic $4$-bracket is generally applicable for many compressible flow models, see \cite{PhysRevE.109.045202, zaidni2024metriplectic}, and directly connects with standard arguments from non-equilibrium thermodynamics \cite{de2013non}, e.g.\ the force and flux pairs from Onsager reciprocity. 

The metriplectic $2$-bracket is then defined to be $(F,G)_{H} \coloneq (F, H; G, H)$. The evolution law $\dot{F} = \{F, H\} + (F, S)_{H}$ for arbitrary $F = F[\rho, m, \sigma]$ recovers the full Navier-Stokes-Fourier system. To be explicit, one finds that
\begin{equation}
    \{F, H\}
    =
    - \int_\Omega
    \bigg[
    m \left(
    F_m \partial_x u
    -
    u \partial_x F_m
    \right)
    + 
    \rho \left(
    F_m \partial_x \eta
    -
    m \partial_x F_{\rho}
    \right) 
    +
    \sigma \left(
    F_m \partial_x T
    -
    u \partial_x F_{\sigma}
    \right)
    \bigg]
    \mathsf{d} x \,,
\end{equation}
where we used the fact that $H_m = m/\rho = u$, $H_\sigma = \partial_s U = T$, and 
\begin{equation}
    H_\rho
    \coloneq 
    \eta
    = 
    \frac{m^2}{2\rho^2}
    +
    U
    +
    \rho U_\rho\left( \rho, \frac{\sigma}{\rho} \right)
    -
    \frac{\rho}{\sigma}
    U_s
    \left( \rho, \frac{\sigma}{\rho} \right)
    \,,
\end{equation}
is related to the enthalpy. Combining this with the dissipative vector field implied by equation \eqref{eq:dissipative_flow}, we obtain the weak evolution equations:
\begin{multline} \label{eq:weak_nsf_system}
    \dot{F}
    =
    \{F, H\} + (F, S)_H \\
    =
    - \int_\Omega
    \bigg[
    m \left(
    F_m \partial_x u
    -
    u \partial_x F_m
    \right)
    + 
    \rho \left(
    F_m \partial_x \eta
    -
    m \partial_x F_{\rho}
    \right) 
    +
    \sigma \left(
    F_m \partial_x T
    -
    u \partial_x F_{\sigma}
    \right)
    \bigg]
    \mathsf{d} x \\
    +
    \int_\Omega
    \left(
    - 
    \left[
     (\partial_x F_m) \mu \partial_x u
    +
    (\partial_x F_\sigma) \frac{\kappa}{T} (\partial_x T)
    \right]
    + 
    F_\sigma
    \left[
    \frac{\mu}{T} (\partial_x u)^2 
    +
    \frac{\kappa}{T^2} (\partial_x T)^2
    \right]
    \right)
    \mathsf{d} x \,.
\end{multline}
Integration by parts and some algebraic manipulation recovers the strong evolution equations given in equation \eqref{eq:nsf_system}. However, it is this weak form in equation \eqref{eq:weak_nsf_system} implied by the Hamiltonian and metriplectic structure, and not the evolution equations themselves in equation \eqref{eq:nsf_system}, from which a thermodynamically-consistent finite element method will be derived. 

It is convenient to non-dimensionalize the equations of motion. The viscosity and conductivity coefficients are assumed to be constant. Define the following dimensionless quantities:
\begin{equation}
    \tilde{x} = \frac{x}{L} \,,
    \quad
    \tilde{u} = \frac{u}{V} \,,
    \quad
    \tilde{\rho} = \frac{\rho}{\rho_0} \,,
    \quad
    \tilde{\sigma} = \frac{\sigma}{\rho_0 R} \,,
    \quad 
    \tilde{t} = \frac{L t}{V} \,,
    \quad
    \tilde{p} = \frac{p}{\rho_0 V^2} \,,
    \quad \text{and} \quad
    \tilde{T} = \frac{R T}{\rho_0 V^2} \,,
\end{equation}
where tildes indicate dimensionless quantities; $L$, $V$, and $\rho_0$ are taken to be the characteristic length, velocity, and density, respectively; and $R$ is the ideal gas constant. Dropping the tildes for notational ease, the equations of motion become
\begin{equation} \label{eq:nsf_system_nondim}
\begin{aligned}
    &\partial_t \rho
    + \partial_x (\rho u) = 0 \,,
    \quad
    \partial_t (\rho u) 
    +
    \partial_x (\rho u^2)
    + \partial_x p
    = \frac{1}{\text{Re}} \partial_x^2 u \,,
    \\
    &\partial_t (\rho s)
    + 
    \partial_x(\rho s u)
    =
    \frac{1}{\text{Re}} \frac{(\partial_x u)^2}{T} 
    + \frac{1}{\text{Re} \, \text{Pr}} \frac{\gamma}{\gamma - 1}
    \left(
    \partial_x \left( \frac{1}{T} \partial_x T \right)
    + \frac{(\partial_x T)^2}{T^2} \right) \,,
\end{aligned}
\end{equation}
where $\text{Re} = \mu/(\rho_0 VL)$, $\text{Pr} = \kappa/(\mu c_P)$, and $\gamma = c_p/c_v$ are the Reynolds number, Prandtl number, and heat capacity ratio respectively. Recall that $R/c_v = \gamma - 1$. 

For the purposes of this paper, it is sufficient to consider the ideal gas equation of state. In dimensionless units, the internal energy is written $U(\rho, s) = \rho^{\gamma - 1} e^{(\gamma - 1) s}$ so that 
$
    p 
    = 
    \rho^2 \partial_1 U 
    = 
    (\gamma - 1 ) \rho^{\gamma} e^{(\gamma - 1) \sigma/\rho}
$, and 
$   T
    =
    \partial_2 U
    =
    (\gamma - 1 ) \rho^{\gamma - 1} e^{(\gamma - 1) \sigma/\rho}
$,
where $\partial_i$ indicates differentiation with respect to the $i^{th}$ argument. In these units $p = \rho T$ as required. 

\section{A thermodynamically consistent discretization}

In this work, we consider simulations on a periodic domain: $\Omega = [0,L]/\!\!\sim$, where $L > 0$ and the equivalence relation identifies the endpoints. Let $V_h \subset H^1(\Omega)$ be the degree-$p$ continuous Galerkin finite element space defined over a uniform grid, $\mathcal{T}_h$, on $\Omega$: i.e.
\begin{equation}
    V_h
    =
    \{ v_h \in H^1(\Omega) : \left. v_h \right|_{K} \in \mathbb{P}^p(K) \,,\, \forall K \in \mathcal{T}_h \} \,,
\end{equation}
where $\mathbb{P}^p(K)$ is the space of degree-$p$ polynomials on $K \subset \Omega$. The discretization is accomplished using the method of lines by positing that all dynamical fields have spatial dependence modeled in this Galerkin subspace. However, rather than discretizing the equations of motion themselves, we discretize the weak forms implied by the metriplectic formulation. 

Let $(\rho_h, m_h, \sigma_h) \in V_h \times V_h \times V_h$. The discretized Hamiltonian and entropy are given by
\begin{equation}
    H^h[\rho_h, m_h, \sigma_h]
    =
    \int_\Omega
    \left[
    \frac{1}{2} \frac{m_h^2}{\rho_h}
    +
    \rho_h U \left(\rho_h, \frac{\sigma_h}{\rho_h} \right) 
    \right]
    \mathsf{d} x \,, 
    \quad
    S^h[\sigma_h]
    =
    \int_\Omega \sigma_h \mathsf{d} x \,,
\end{equation}
the antisymmetric bracket is given by
\begin{multline} \label{eq:discrete_poisson_bracket}
    \{F^h, G^h\}_h(\rho_h, m_h, \sigma_h)
    =
    - \int_\Omega
    \bigg[
    m_h \left(
    F^h_{m_h} \partial_x G^h_{m_h}
    -
    G^h_{m_h} \partial_x F^h_{m_h}
    \right) \\
    + 
    \rho_h \left(
    F^h_{m_h} \partial_x G^h_{\rho_h}
    -
    G^h_{m_h} \partial_x F^h_{\rho_h}
    \right) 
    +
    \sigma_h \left(
    F^h_{m_h} \partial_x G^h_{\sigma_h}
    -
    G^h_{m_h} \partial_x F^h_{\sigma_h}
    \right)
    \bigg]
    \mathsf{d} x \,,
\end{multline}
and the metriplectic $4$-bracket is given by
\begin{multline}
    (F^h, K^h; G^h, N^h)_h 
    = 
    \frac{1}{\text{Re}}
    \int_\Omega
    \frac{1}{T_h}
    \bigg[
    \left(
    K^h_{\sigma_h} \partial_x F^h_{m_h}
    -
    F^h_{\sigma_h} \partial_x K^h_{m_h}
    \right) 
    \left(
    N^h_{\sigma_h} \partial_x G^h_{m_h}
    -
    G^h_{\sigma_h} \partial_x N^h_{m_h}
    \right) \\
    +
    \frac{1}{\text{Pr}} \frac{\gamma}{\gamma - 1}
    \frac{1}{T_h}
    \left(
    K^h_{\sigma_h} \partial_x F^h_{\sigma_h}
    -
    F^h_{\sigma_h} \partial_x K^h_{\sigma_h}
    \right) 
    \left(
    N^h_{\sigma_h} \partial_x G^h_{\sigma_h}
    -
    G^h_{\sigma_h} \partial_x N^h_{\sigma_h}
    \right)
    \bigg]
    \mathsf{d} x \,,
\end{multline}
where $F^h = \left. F \right|_{V_h}$, and similarly for the other functionals. We call the bracket in equation \eqref{eq:discrete_poisson_bracket} an antisymmetric bracket, and not a Poisson bracket, because it is fails to satisfy the Jacobi identity: i.e. the identity
\begin{equation}
    \{ F, \{G, H\}\}
    +
    \{ H, \{F, G\}\}
    +
    \{ G, \{H, F\}\}
    = 0 \quad \forall F, G, H \,.
\end{equation}
This is an essential algebraic property of Poisson brackets. However, no grid-based discretization of the kinds of Poisson brackets found in fluid models (or indeed those of most Hamiltonian partial differential equations) which preserves the Jacobi identity is known. This deficiency motivates the use of the terminology ``almost Poisson" sometimes found in the literature \cite{cotter2023compatible} to describe discretizations of Poisson brackets which fail to satisfy the Jacobi identity. These discretizations nonetheless preserve antisymmetry and the Casimir invariants giving rise to mass and total entropy conservation, which is sufficient for the purposes of this work. 

The functional derivatives of the Hamiltonian are as follows: 
\begin{multline}
    H_{\rho_h}^h
    =
    Q_{V_h} \bigg( 
    - \frac{m_h^2}{2 \rho_h^2} 
    + U \left(\rho_h, \frac{\sigma_h}{\rho_h} \right) 
    + \rho_h \partial_1 U \left(\rho_h, \frac{\sigma_h}{\rho_h} \right) 
    - \frac{\rho_h}{\sigma_h} \partial_2 U  \left(\rho_h, \frac{\sigma_h}{\rho_h} \right)
    \bigg) \,, \\
    H_{m_h}^h
    =
    Q_{V_h} \left( \frac{m_h}{\rho_h} \right) \,, 
    \quad \text{and} \quad
    H_{\sigma_h}^h
    =
    Q_{V_h} \left( \partial_2 U  \left(\rho_h, \frac{\sigma_h}{\rho_h} \right) \right) \,,
    \quad
\end{multline}
where $Q_{V_h}$ is the $L^2$ projection onto $V_h$. These derivatives must be projected because the functional derivatives are taken with respect to constrained variations in the space $V_h$. Similarly, one finds $S_{\rho_h}^h = S_{m_h}^h = 0$, and $S_{\sigma_h}^h = 1$, since $V_h$ interpolates constant functions exactly. For convenience, and to match notation used subsequently, we write $\delta \bm{s}_h = (0, 0, 1)$ to denote the vector of derivatives of the entropy with respect to the three dynamical fields, $(\rho_h, m_h, \sigma_h)$. 

The evolution is then given by $\dot{F}^h = \{F^h, H^h\}_h + (F^h, H^h; S^h, H^h)_h$. One immediately finds that the semi-discrete model is thermodynamically consistent, $\dot{H}^h = 0$ and $\dot{S}^h \geq 0$, as the discretized brackets possess the same symmetries and degeneracies as the continuous brackets. If we consider an observable of the form $F^h = (\phi_m, m_h)_{L^2} + (\phi_\rho, \rho_h)_{L^2} + (\phi_\sigma, \sigma_h)_{L^2}$, then we obtain the following variational problem: find $(\bm{u}_h, \delta \bm{h}_h) \coloneq ((\rho_h, m_h, \sigma_h), (\eta_h, u_h, T_h)) \in V_h^3 \times V_h^3$, where $V_h^3 = V_h \times V_h \times V_h$, such that
\begin{equation} \label{eq:evolution_eqns_fem_model}
    (\bm{v}_h, \partial_t \bm{u}_h)_{L^2}
    -
    \{ \bm{v}_h, \delta \bm{h}_h \} \left( \bm{u}_h \right) 
    -
    ( \bm{v}_h, \delta \bm{s}_h)_H ( \delta \bm{h}_h ) \\
    +
    (\delta \bm{h}_h - DH(\bm{u}_h), \bm{w}_h)_{L^2}
    =
    0 
\end{equation}
$\forall (\bm{v}_h, \bm{w}_h) \coloneq ((\phi_\rho, \phi_m, \phi_\sigma), (\phi_{\eta_h}, \phi_{u_h}, \phi_{T_h})) \in V_h^3 \times V_h^3$, where 
\begin{equation}
    (\bm{v}_h, \partial_t \bm{u}_h)_{L^2}
    =
    (\partial_t \rho_h, \phi_\rho)_{L^2}
    +
    (\partial_t m_h, \phi_m)_{L^2}
    +
    (\partial_t \sigma_h, \phi_\sigma)_{L^2} \,,
\end{equation}
the discrete Poisson bracket is defined to be
\begin{multline}
    \{ \bm{v}_h, \delta \bm{h}_h \} \left( \bm{u}_h \right)
    =
    \{ F^h, H^h \}_h(\bm{u}_h) 
    =
    - \left( m_h \partial_x u_h, \phi_m \right)_{L^2}
    +
    \left( m_h u_h, \partial_x \phi_m \right)_{L^2} \\
    -
    \left( \rho_h \partial_x \eta_h, \phi_m \right)_{L^2} 
    +
    \left( \rho_h u_h, \partial_x \phi_\rho \right)_{L^2} 
    -
    \left( \sigma_h \partial_x T_h, \phi_m \right)_{L^2}
    +
    \left( \sigma_h u_h, \partial_x \phi_\sigma \right)_{L^2} \,,
\end{multline}
the discrete metriplectic bracket yields
\begin{multline}
    (\bm{v}_h, \delta \bm{s}_h)_H ( \delta \bm{h}_h )
    =
    (F^h, H^h; S^h, H^h)_h 
    =
    - \frac{1}{\text{Re}}
    \Bigg[
    \left(\partial_x u_h, \partial_x \phi_m \right)_{L^2} 
    -
    \left(
    \frac{ (\partial_x u_h)^2}{T_h}, \phi_\sigma
    \right)_{L^2} \\
    +
    \frac{1}{\text{Pr}} \frac{\gamma}{\gamma - 1}
    \bigg[
    \left(
    \frac{\partial_x T_h}{T_h}, \partial_x \phi_\sigma
    \right)_{L^2}     
    -
    \left(
    \frac{ (\partial_x T_h)^2}{T_h^2}, \phi_\sigma
    \right)_{L^2}
    \bigg]
    \Bigg] \,,
\end{multline}
and the $L^2$ projections of the derivatives of the Hamiltonian are imposed via
\begin{equation}
\begin{aligned}
    (\delta \bm{h}_h - DH(\bm{u}_h), \bm{w}_h)_{L^2}
    =
    \left(
    \eta_h
    -
    \frac{\delta H^h}{\delta \rho_h},
    \phi_{\eta_h}
    \right)_{L^2}
    + 
    \left( 
    u_h - \frac{\delta H^h}{\delta m_h}, 
    \phi_{u_h} 
    \right)_{L^2} 
    +
    \left( 
    T_h - \frac{\delta H^h}{\delta \sigma_h},
    \phi_{T_h}
    \right)_{L^2} \,.
\end{aligned}
\end{equation}
This notation attempts to stress three essential features of the spatial discretization.
\begin{itemize}
    \item The derivatives of the generating functions are computed as via projections and must be thought of as distinct from the evolving state vector, $\bm{u}_h = (\rho_h, m_h, \sigma_h)$. Hence, we keep track of an the derivatives of the Hamiltonian, $\delta \bm{h}_h = (\eta_h, m_h, T_h)$, as additional degrees of freedom (note, $\delta \bm{s}_h = (0, 0, 1)$ takes a simple form in momentum coordinates).
    \item The bilinear two-brackets generating the conservative and dissipative dynamics, 
    \begin{equation}
        \{ \bm{v}_h, \delta \bm{h}_h \} \left( \bm{u}_h \right)
        \quad \text{and} \quad
        (\bm{v}_h, \delta \bm{s}_h)_H ( \delta \bm{h}_h ) \,,
    \end{equation}
    respectively, take the derivatives of the generating functions, $\delta \bm{h}_h$ and $\delta \bm{s}_h$, as one argument, and arbitrary the test function, $\bm{v}_h$ as the other. 
    \item These brackets also have nonlinear field dependence. The discrete antisymmetric bracket depends directly on the state-vector, $\bm{u}_h$, while the discrete symmetric bracket depends on the derivative of the Hamiltonian, $\delta \bm{h}_h$. The nonlinear dependence of the dissipative bracket on $\delta \bm{h}_h$, rather than $\bm{u}_h$, is dictated by the $4$-bracket formalism and essential for energy conservation. 
\end{itemize}

By including the $L^2$ projection of the derivatives of the Hamiltonian as additional fields to solve for in the variational problem, we formulate the semi-discrete problem as a differential algebraic equation. The derivatives of the Hamiltonian with respect to momentum and entropy density have the physical interpretation of being the velocity and temperature, respectively. The derivative of the Hamiltonian with respect to density is related to the enthalpy, and one may readily recover the gradient of the pressure through a Bernoulli-like equation:
\begin{equation} \label{eq:proj}
    \left( \partial_x p_h + \frac{1}{2} \partial_x (u_h^2) - \rho_h \partial_x \eta_h + \sigma_h \partial_x T_h, \phi
    \right)_{L^2} = 0 \,, \quad \forall \phi \in V_h \,.
\end{equation}

To be perfectly explicit, the variational form for the momentum equation is obtained as follows. Letting $\phi_\rho = \phi_\sigma = 0$, we find
\begin{multline}
    (\phi_m, \partial_t m_h)
    + \left( m_h \partial_x u_h, \phi_m \right)_{L^2}
    -
    \left( m_h u_h, \partial_x \phi_m \right)_{L^2} \\
    +
    \left( \rho_h \partial_x \eta_h, \phi_m \right)_{L^2} 
    +
    \left( \sigma_h \partial_x T_h, \phi_m \right)_{L^2} 
    +
    \frac{1}{\text{Re}}
    \left(\partial_x u_h, \partial_x \phi_m \right)_{L^2} 
    =
    0 \,, \quad \forall \phi_m \in V_h \,,
\end{multline}
where
\begin{multline}
    \bigg( 
    \eta_h
    + \frac{m_h^2}{2 \rho_h^2} 
    - U \left(\rho_h, \frac{\sigma_h}{\rho_h} \right) 
    - \rho_h \partial_1 U \left(\rho_h, \frac{\sigma_h}{\rho_h} \right) 
    + \frac{\rho_h}{\sigma_h} \partial_2 U  \left(\rho_h, \frac{\sigma_h}{\rho_h} \right) ,
    \phi_\eta
    \bigg)_{L^2} = 0 \,, \\
    \left(u_h - \frac{m_h}{\rho_h}, \phi_u \right)_{L^2} = 0 \,, 
    \quad \text{and} \quad
    \left(T_h - \partial_2 U  \left(\rho_h, \frac{\sigma_h}{\rho_h} \right), \phi_T \right)_{L^2} = 0 \,, \quad \forall (\phi_\eta, \phi_u, \phi_T) \in V_h^3 \,.
\end{multline}
The continuity and entropy equations are obtained in a similar fashion. 

As mentioned previously, the spatially semi-discretized evolution equations given in equation \eqref{eq:evolution_eqns_fem_model} are thermodynamically consistent. This may be verified by letting $\bm{v}_h = \delta \bm{h}_h$, yielding
\begin{equation}
    \dot{H}^h
    =
    (\delta \bm{h}_h, \partial_t \bm{u}_h)_{L^2}
    =
    0 \,,
\end{equation}
and $\bm{v}_h = \delta \bm{s}_h$, yielding
\begin{equation}
    \dot{S}^h
    =
    (\delta \bm{s}_h, \partial_t \bm{u}_h)_{L^2}
    =
    \frac{1}{\text{Re}}
    \Bigg[
    \left(
    \frac{ (\partial_x u_h)^2}{T_h}, 1
    \right)_{L^2} 
    +
    \frac{1}{\text{Pr}} \frac{\gamma}{\gamma - 1}
    \left(
    \frac{ (\partial_x T_h)^2}{T_h^2}, 1
    \right)_{L^2}
    \Bigg] \geq 0 \,.
\end{equation}

\section{Temporal discretization}

One convenient and simple choice for temporal discretization is the implicit midpoint method. That is, for a differential equation $\dot{z} = V(z)$, its evolution is given by
\begin{equation}
    \frac{z^{n+1} - z^n}{\Delta t}
    =
    V
    \left(
    \frac{z^{n+1} + z^n}{2}
    \right) \,.
\end{equation}
This is done because the method is symplectic, A-stable, and known to preserve invariants well: quadratic invariants are preserved exactly \cite{hairer2013geometric}. Mass is conserved exactly, and in the dissipation-free limit, so is entropy. The energy is not a polynomial invariant and therefore is not conserved exactly even in the dissipation-free limit. In fact, as the spatially semi-discrete model is not Hamiltonian even in the dissipation-free limit (although it does conserve energy due to antisymmetry of the Poisson bracket and degeneracy of the metriplectic bracket) there is no guarantee of the long-time near energy conservation property symplectic integrators usually enjoy \cite{hairer2013geometric}. This is because the proof of long-time energy conservation for symplectic integrators applied to Hamiltonian systems crucially relies on the Hamiltonian structure, namely that the time-advance map is a canonical transformation. In fact, a small drift in energy is observed in the numerical results section in both the dissipation-free and dissipative test cases, see Figures \ref{subfig:cons_laws_cons} and \ref{subfig:cons_laws_diss} respectively. Entropy production of the fully discrete system is given by
\begin{equation}
    \frac{S^{n+1} - S^n}{\Delta t} 
    = \frac{1}{\text{Re}}
    \left[
    \left(
    \frac{ (\partial_x u_h^n)^2}{T_h^n}, 
    1
    \right)_{L^2} 
    +
    \frac{1}{\text{Pr}} \frac{\gamma}{\gamma - 1}
    \left(
    \frac{ (\partial_x T_h^n)^2}{(T_h^n)^2}, 
    1
    \right)_{L^2}
    \right]
    \geq 0 \,.
\end{equation}

The failure of the implicit midpoint method to yield a thermodynamically-consistent time-discretization motivates us to consider a time-stepping strategy based on the averaged vector-field discrete gradient method \cite{quispel2008new, hairer2010energy}. The time-stepping method based on the averaged vector-field discrete gradient method for equation \eqref{eq:evolution_eqns_fem_model} is given by the weak form
\begin{multline}
    \left(\frac{\bm{u}_h^{n+1} - \bm{u}_h^n}{\Delta t}, \bm{v}_h \right)_{L^2}
    +
    \{ \delta \bm{h}_h^n, \bm{v}_h \} \left( \frac{\bm{u}_h^{n+1} + \bm{u}_h^n}{2} \right) 
    + 
    ( \delta \bm{s}_h^n, \bm{v}_h)_H ( \delta \bm{h}_h^n ) \\
    +
    ( \overline{DH}(\bm{u}_h^{n}, \bm{u}_h^{n+1}) - \delta \bm{h}_h^n, \bm{w}_h)_{L^2}
    =
    0 \,, \quad \forall (\bm{v}_h, \bm{w}_h) \in V_h^3 \times V_h^3
\end{multline}
where
\begin{equation} \label{eq:avg_vector_field_hamiltonian}
    \overline{DH}(\bm{u}_h^{n}, \bm{u}_h^{n+1})
    =
    \int_0^1 DH( (1-t) \bm{u}_h^n + t \bm{u}_h^{n+1}) \mathsf{d} t \,.
\end{equation}
This method is equivalent to the implicit midpoint method if we approximate the integral in \eqref{eq:avg_vector_field_hamiltonian} using the midpoint rule. In fact, this integral must be approximated via quadrature in general. We find that Gauss-Legendre quadrature with $\geq 4$ quadrature points achieves sufficient accuracy to achieve energy conservation to machine precision in the tests considered in this work. From this definition of the time-stepping scheme, it follows that if we let $\bm{v}_h = \delta \bm{h}_h$, then the fundamental theorem of calculus implies that
\begin{equation}
\begin{aligned}
    \left(\frac{\bm{u}_h^{n+1} - \bm{u}_h^n}{\Delta t}, \delta \bm{h}_h \right)_{L^2}
    &=
    \left(\frac{\bm{u}_h^{n+1} - \bm{u}_h^n}{\Delta t}, \overline{DH}(\bm{u}_h^{n}, \bm{u}_h^{n+1}) \right)_{L^2} \\
    &=
    \frac{1}{\Delta t} \int_0^1
    DH( (1-t) \bm{u}_h^n + t \bm{u}_h^{n+1})
    \cdot (\bm{u}_h^{n+1} - \bm{u}_h^n) 
     \mathsf{d} t
    \\
    &=
    \frac{1}{\Delta t} \int_0^1
    \frac{\mathsf{d}}{\mathsf{d} t} H ((1 - t) \bm{u}_h^{n} + t \bm{u}_h^{n+1}) \mathsf{d} t
    =
    \frac{H(\bm{u}_h^{n+1}) - H(\bm{u}_h^n)}{\Delta t}
    =
    0 \,,
\end{aligned}
\end{equation}
verifying energy conservation. Positive entropy production follows from letting $\bm{v}_h = \delta \bm{s}_h^n$: 
\begin{equation}
    \frac{S^{n+1} - S^n}{\Delta t} 
    =
    ( \delta \bm{s}_h^n, \delta \bm{s}_h^n)_H ( \delta \bm{h}_h )
    = \frac{1}{\text{Re}}
    \left[
    \left(
    \frac{ (\partial_x u_h^n)^2}{T_h^n}, 
    1
    \right)_{L^2} 
    +
    \frac{1}{\text{Pr}} \frac{\gamma}{\gamma - 1}
    \left(
    \frac{ (\partial_x T_h^n)^2}{(T_h^n)^2}, 
    1
    \right)_{L^2}
    \right]
    \geq 0 \,.
\end{equation}
Hence, the fully discrete method is found to be thermodynamically-consistent. This is verified in figures \ref{subfig:cons_laws_disc_grad_cons} and \ref{subfig:cons_laws_disc_grad_diss} for both the dissipation-free and dissipative test cases. The averaged vector field discrete gradient method is $O(\Delta t^2)$, however higher order generalizations were derived in \cite{cohen2011linear}. Moreover, both the Gauss-Legendre implicit Runge-Kutta methods and energy conserving methods of the kind found in \cite{cohen2011linear} were recently shown to fit into a general framework in \cite{andrews2024high}. 

\section{Numerical examples}

\begin{figure}[ht]
\centering
\subfigure[$\text{Re} = 10$]{\includegraphics[width=0.95\linewidth]{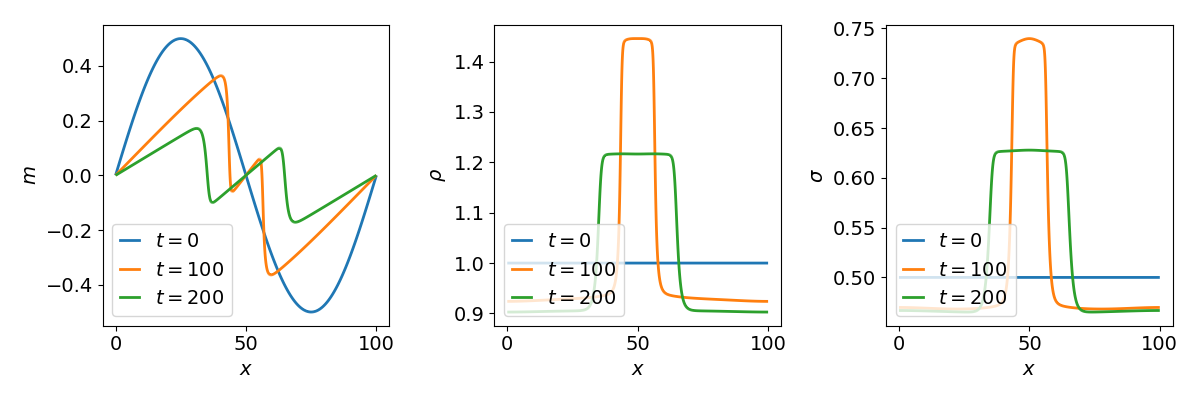}}
\subfigure[$\text{Re} = \infty$]{\includegraphics[width=0.95\linewidth]{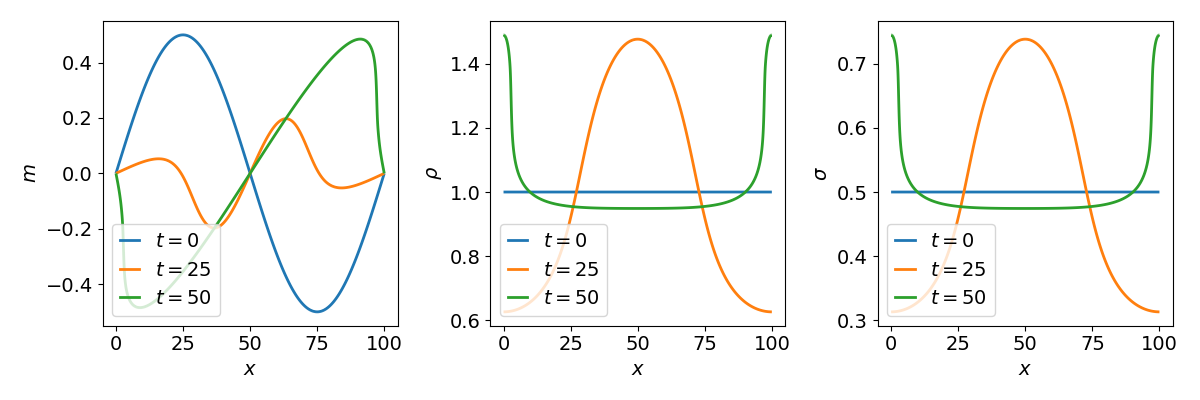}}
\caption{Visualization of solutions. The dissipation-free solution is shown prior to shock formation at $t \approx 50$. The results using implicit midpoint and discrete-gradient time-stepping look indistinguishable to the eye.}
\label{fig:results}
\end{figure}

\begin{figure}[ht]
    \centering
    \subfigure[$\text{Re} = \infty$ with implicit midpoint time-stepping.\label{subfig:cons_laws_cons}]{
        \includegraphics[width=0.45\linewidth]{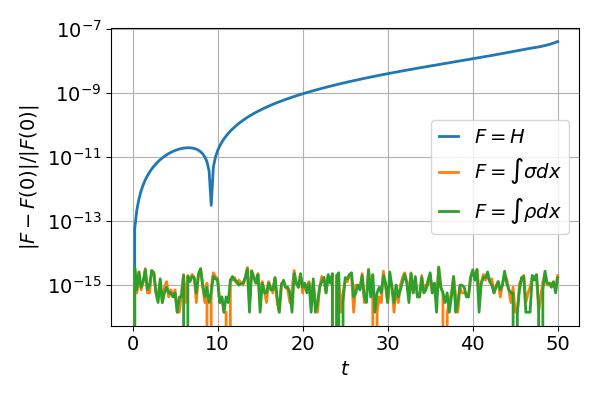}
    }
    \subfigure[$\text{Re} = 10$ with implicit midpoint time-stepping.\label{subfig:cons_laws_diss}]{
        \includegraphics[width=0.45\linewidth]{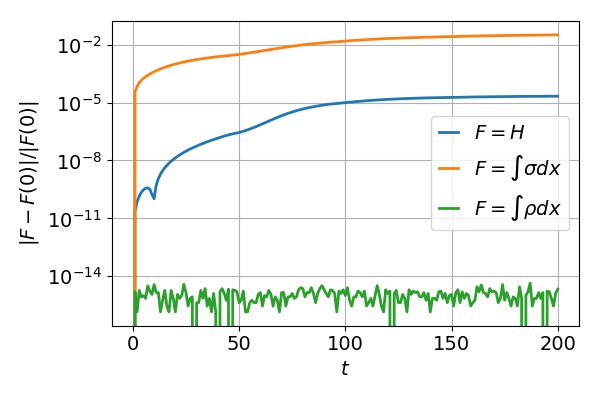}
    }
    \subfigure[$\text{Re} = \infty$ with discrete gradient time-stepping.\label{subfig:cons_laws_disc_grad_cons}]{
        \includegraphics[width=0.45\linewidth]{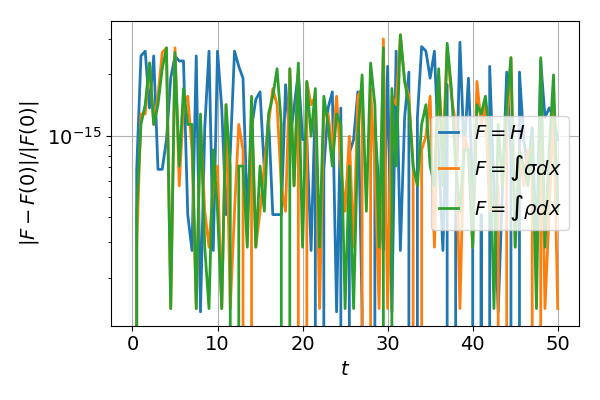}
    }
    \subfigure[$\text{Re} = 10$ with discrete gradient time-stepping.\label{subfig:cons_laws_disc_grad_diss}]{
        \includegraphics[width=0.45\linewidth]{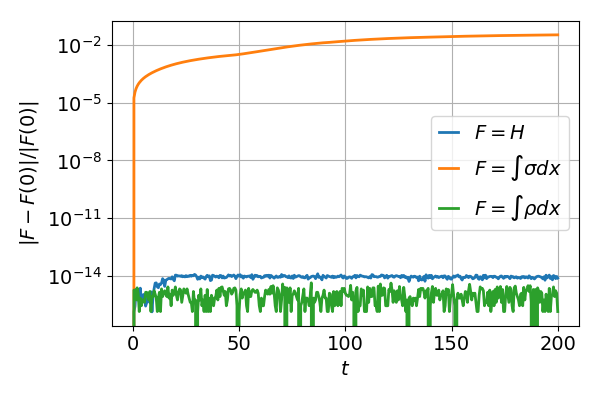}
    }
    \caption{Evolution of relative error in total energy, entropy, and mass.}
    \label{fig:cons_laws}
\end{figure}

The spatial discretization is accomplished using the Firedrake library \cite{FiredrakeUserManual}, and the temporal discretization with the Irksome module \cite{10.1145/3466168}. For the finite element discretizations, we use piecewise linear interpolation. Although there is no inherent limitation which forces one to use linear finite elements, sharp gradients form in this compressible flow problem making it advantageous to use a fine grid with low order interpolation. In the following examples, we use the parameters
\begin{equation} \label{eq:params}
    \text{Re} = 10 \,,
    \quad
    \text{Pr} = 0.71 \,,
    \quad \text{and} \quad
    \gamma = 1.4 
\end{equation}
to reflect the standard parameters of dry air with a relatively low Reynolds number (so that the effects of dissipation might be readily seen). We present two simulations with initial conditions $m_h(x, 0) = \sin(2 \pi x/L)/2$, $\rho_h(x,0) = 1$, and $\sigma_h(x,0) = 1/2$. In one simulation, we use the parameter set \eqref{eq:params}, in the other, we let $\text{Re} \to \infty$ to simulate the dissipation-free dynamics (terminated prior to shock formation). The spatial domain is taken to be $[0,L] = [0, 100]$. Tests are run using both the implicit midpoint and discrete gradient time-stepping schemes. In all tests, the time step is taken to be $\Delta t = 0.1$, and the grid size is $\Delta x = L/2000 = 0.05$. For the dissipative simulations, the simulation is run for $t \in [0,200]$, while the dissipation-free simulation is run for $t \in [0,50]$ (due to the lack of viscous regularization, a shock forms at $t\approx50$). See Figure \ref{fig:results} for a visualization of the simulation results, and Figure \ref{fig:cons_laws} for a visualization of the mass, energy, and entropy as a function of time for each simulation. 

As previously mentioned, the implicit midpoint method fails to conserve energy whereas the discrete gradient method does, as seen in Figure \ref{fig:cons_laws}. Because the dissipation-free system is not Hamiltonian, there is no guarantee that a symplectic integrator should enjoy long-time energy conservation. However, even if the dissipation-free spatially discrete system were Hamiltonian---so that symplectic integration yielded a long-time energy near-conservation result---proving energy conservation for the fully-discrete metriplectic system remains problematic. The conserved modified energy obtained through backward error analysis would most likely fail to lie in the null space of the metriplectic bracket. Thus, overall energy conservation of the coupled conservative-dissipative dynamics remains uncertain even in this optimistic case. For these reasons, symplectic integration is not an appropriate choice for the time-integration of metriplectic systems. Rather, energy conserving methods---such as the averaged vector field discrete gradient method used in this work---are more appropriate. 

\section{Conclusion}

In this work, we derived a thermodynamically consistent discretization of the one dimensional Navier-Stokes-Fourier model using the metriplectic $4$-bracket formalism. For Galerkin methods, one simply restricts the brackets and functionals to act on finite dimensional function spaces. A comparable discretization using finite-differences could be derived by directly approximating the functionals and brackets using quadrature. Virtually any spatial discretization method, if applied at the level of the brackets and generating functions, would yield a thermodynamically consistent spatial semi-discretization as long as the resulting discrete brackets retain the symmetries and degeneracies of the continuous formulation. Many other models fit into the metriplectic formalism \cite{PhysRevE.109.045202, osti_2426411, SATO2024100054}, and one may reasonably expect their discretization to likewise be thermodynamically consistent if one uses analogous methods to those employed in this paper. The implicit midpoint method was found to yield largely favorable behavior as a time-integrator, however it fails to exactly conserve energy. The averaged vector field discrete gradient method \cite{hairer2010energy, mclachlan1999geometric} applied to the bracket-based spatial discretization yielded a thermodynamically consistent fully-discrete method. Other energy-conserving methods such as those found in \cite{cohen2011linear, andrews2024high} could likewise be used. The simultaneous guarantee of energy conservation and the positive production of physical entropy at the correct rate makes these energy-conserving methods preferable to standard time-stepping methods. Finally, compressible flow models exhibit discontinuous shock solutions in the inviscid case, and effective shocks if the spatial discretization does not resolve the viscous boundary layer. A thermodynamically consistent spatial discretization based on a discontinuous Galerkin, finite volume, or finite difference method with stabilization for shock solutions using the metriplectic formalism is an intriguing future direction of inquiry. 

\section*{Acknowledgements}

A.Z. acknowledges support from the Mohammed VI Polytechnic University. W.B. and P.J.M. acknowledge support from the DOE Office of Fusion Energy Sciences under DE-FG02-04ER-54742. We also thank Chris Eldred for insightful discussions on this work. 



\bibliographystyle{elsarticle-num} 
\bibliography{references}





\end{document}